\documentclass[journal]{IEEEtran}
\usepackage{algorithm}
\usepackage{algorithmic}
\usepackage{multirow}
\usepackage{epsfig}
\usepackage{amssymb}
\usepackage{amsmath}
\usepackage{amsfonts}

\usepackage{subfig}             
\usepackage{array}
\usepackage{epsf}
\usepackage{epstopdf}
\usepackage{lineno,hyperref}
\modulolinenumbers[5]
\usepackage{graphicx}
\usepackage{threeparttable}
\usepackage[draft]{todonotes}   

\begin{document}

\title{Implicit Multi-feature Learning for Dynamic Time Series Prediction of the Impact of Institutions}

\author{Xiaomei Bai, Fuli Zhang, Jie Hou, Feng Xia, Amr Tolba, and Elsayed Elashkar

\thanks{The authors extend their appreciation to the Deanship of Scientific Research at King Saud University for funding this work through research group NO (RGP-1438-27).}

\thanks{X. Bai, J. Hou, and F. Xia are with the Key Laboratory for Ubiquitous Network and Service Software of Liaoning Province, School of Software, Dalian University of Technology, Dalian 116620, China.}
\thanks{X. Bai is also with Computing Center, Anshan Normal University, Anshan 114007, China.}
\thanks{F. Zhang is Library, Anshan Normal University, Anshan 114007, China.}
\thanks{A. Tolba is jointly with Computer Science Department, Community College, King Saud University, Riyadh 11437, Saudi Arabia and Mathematics Department, Faculty of Science, Menoufia University, Shebin El-Kom 32511, Egypt}
\thanks{E. Elashkar is jointly with Administrative Sciences Department, Community College, King Saud University, Riyadh 11437, Saudi Arabia and Applied Statistics Department, Faculty of Commerce, Mansoura University, Egypt.}

\thanks{Corresponding Author: Feng Xia; Email: f.xia@ieee.org.}

}
\markboth{IEEE Access}%
{Shell \MakeLowercase{\textit{et al.}}: Bare Demo of IEEEtran.cls for IEEE Journals}

\maketitle

\begin{abstract}
Predicting the impact of research institutions is an important tool for decision makers, such as resource allocation for funding bodies. Despite significant effort of adopting quantitative indicators to measure the impact of research institutions, little is known that how the impact of institutions evolves in time. Previous researches have focused on using the historical relevance scores of different institutions to predict potential future impact for these institutions. In this paper, we explore the factors that can drive the changes of the impact of institutions, finding that the impact of an institution, as measured by the number of the accepted papers of the institution, more is determined by the authors' influence of the institution. Geographic location of institution feature and state GDP can drive the changes of the impact of institutions. Identifying these features allows us to formulate a predictive model that integrates the effects of individual ability, location of institution, and state GDP. The model unveils the underlying factors driving the future impact of institutions, which can be used to accurately predict the future impact of institutions.
\end{abstract}

%
%
\section{Introduction}
\IEEEPARstart{W}{ith} the rapid growth of scholarly big data~\cite{Xia2017Big,Wang2017Scientific}, the scientific future impact often plays an important role, which can help decision-makers to make better decisions. For example, the prediction of the impact of papers, scholars, and institutions can guide government funding allocation and assess the grant proposals.

The evaluation and prediction of scholarly impact are two important aspects of scientific impact~\cite{Setti2013Bibliometric}. Scholarly impact evaluation focuses on quantifying the previous impact of scholarly entities~\cite{Petersen2014Reputation,Xia2016Bibliographic,Bai2016Identifying}. The institution impact evaluation mainly has two categories: full counting and fractional counting. The former considered that all authors had the same contributions for a paper~\cite{bensman2011evaluation,vinkler2010evaluation}. The latter considered the best journal and highly-cited papers~\cite{bornmann2014mapping}. Due to lack the baseline, the prediction of institution impact is challenging. The KDD CUP 2016 offers a evaluation criterion relying on the number of accepted papers. Previous researchers mainly used the historical relevance scores of each institution to predict the impact of the institution in next year. Based on the gradient boost decision trees (GBDT) model~\cite{Hastie2009Boosting}, Sandulescu et al.~\cite{sandulescu2016predicting} leveraged the following features: the historical relevance scores of the accepted papers of every institution, Author Impact Factor (AIF)~\cite{Pan2014Author}, and the weighted moving-average of relevance scores from previous years to predict the accepted papers of the institution in next year. Xie~\cite{Xie2016Predicting} used linear regression and gradient boosting decision tree to predict the impact of each institution by integrating four features: accepted paper-rank, program committee membership, cross-conference, and cross-phase. Orouskhani et al.~\cite{Orouskhani2016Ranking} ranked the research institutions based on the annual scores of the institutions.

Compared to the evaluation of scholarly impact, to find future potential impact is more guiding significance. Exploring the factors driving scholarly impact is very crucial. Predicting the impact of papers usually focused on the citations prediction or citation distributions. Early citations as a crucial feature usually were used to predict the potential citations of a paper~\cite{bruns2015research,Klimek2016Successful,Cao2016A}. Cao et al.~\cite{Cao2016A} predicted future citations of a paper by integrating a short-term citation history to a Gaussian mixture model. Stegehuis et al.~\cite{Stegehuis2015Predicting} predicted the citations distribution of a paper in the future by the two crucial features: early citations and Journal Impact Factor~\cite{Bornmann2016The}. The prediction of scholars' impact mainly focused on predicting the author's H-index and his (her) citations. Penner et al.~\cite{Penner2013On} applied the future impact model based on linear regression to 762 careers from three disciplines: mathematics, physics and biology. They found that their models' prediction performance depended heavily on scientists' career age. Based on linear regression, Dong et al.~\cite{Dong2016Can} leveraged six factors, including author, content, venue, social information, reference and temporal data to predict an author's H-index~\cite{hirsch2005index} in five years. They found that topic authority and venue were two crucial factors, which can determine whether a newly published paper will enhance its authors' H-index. Based on a linear regression with elastic net regularization, Acuna et al.~\cite{Acuna2012Future} constructed a simple model including the number of papers, H-index, years since first publication, and etc.

Intuitively, the previous changing trend of the impact of each institution is the most relevant to future impact of the institution. Therefore, previous scholars mainly consider the historical relevance scores of the accepted papers for each institution to predict future impact of the institution. Namely, the number of the accepted papers for each institution in previous years is used as an important feature to construct the predictive model. However, we find that scholars' impact such as AIF, Q value~\cite{Sinatra2016Quantifying}, and H-index are more relevant to predict the impact of institutions for top conferences, compared to the relevance scores of the institution. Svider et al.~\cite{Svider2016Are} found that there was an association between industry support and academic impact. An interesting phenomenon is that industry payments greater than $\$10,000$ were related to a greater scholarly impact. Their work inspires us to explore GDP to improve the performance of the predictive model for accurately predicting the potential impact of institutions.

The paper aims to explore the factors driving the changes of the impact of institutions and the contributions of these factors for predicting the impact of the institutions. Via feature selection, we find that the impact of an institution, as measured by the number of the accepted papers of the institution, more is determined by the authors' influence of the institution. Geographic location of institution feature and state GDP can drive the changes of the impact of institutions. At the same time, we also find that the features driving the changing of the impact of institution play different roles for predicting the future potential impact. Based on XGBoost ~\cite{chen2016xgboost}, we propose a novel prediction model, which has the ability to generate accurate predictions and explain the prediction performance.

\section{Methods}
We now describe our methods on predicting the number of the accepted papers of each institution for top conferences, including the following five parts: prediction task, dataset and data processing, factors that drive the impact of institutions to increase, feature selection, and predicting the future impact of institutions.
\subsection{Prediction task}
Given the heterogeneous characteristics of scholarly data, our task aims to predict the number of the accepted papers of each institution for top conferences. We consider the problem of the impact of institutions prediction from academic data. Let Y=\{$Y_{1},Y_{2},\cdots,Y_{T}$\} be the set of the number of the accepted papers for an institution of a conference in different years, where $Y_{i}$ corresponds to the number of the accepted papers for an institution in the $i^{th}$ year. Given features $X$ extracted from the experimental data as input, the impact of institution predictor needs to generate $f=(f_{1},f_{2},\cdots,f_{T})$ as output, where $f_{i}$ is the predicted number of the accepted papers of an institution for a top conference in next year.
\subsection{Dataset and data processing}
We use the real-world data from Microsoft Academic Graph (MAG), which is a large and heterogeneous graph. Each paper contains publication date, citation relationships, authors, institutions, journal or conference, and fields of study. In our experiments, we follow KDD-CUP 2016's conference selection, including FSE, ICML, KDD, MM, MobiCom, SIGCOMM, SIGIR and SIGMOD to construct our experimental dataset. Firstly, we retain the data from 2000 to 2015, and add the loss authors' institution information according to the institution information of their previous publication. Secondly, because a small part of the data is incomplete, the data are deleted. Finally, we remove the duplicated institutional information to find the coordinates of these institutions. After the data processing, our experimental dataset includes 33,953 authors with 19,343 papers from 4,524 institutions across years 2000 to 2015.
\subsection{Factors that drive the impact of institutions to increase}
\textbf{\emph{Author-based features}}.
\begin{itemize}
  \item Author Impact Factor (AIF). \\
  Author impact factor is an extension of the journal impact factor to authors. AIF of an author in year $T$ is the average citations of published papers in a period of $\Delta T$ years before year $T$.
  Based on the eight top conferences selected by KDD-CUP 2016, we compute each author's AIF value according to the author's publishing history
 and use the statistics of a given institution's all authors' AIF as a group of its features, including sum, maximum, minimum, median, average and deviation. We briefly explore and report the authors' AIF features in this work.
  \item $Q$ value.\\
The $Q$ value reflects the ability of scientists to enhance the impact of a paper~\cite{Sinatra2016Quantifying}, and it is a constant in a scientist's career. 
\begin{equation}\label{eq:e1}
Q_{i}=e^{ \left \langle  log c_{i\alpha}\right \rangle-\mu_{p}}
\end{equation}
where $Q_{i}$ indicates a scientist $i$'s $Q$ value. $\left \langle  log c_{i\alpha}\right \rangle$ is the average logarithmic citations of all papers published by scientist $i$ and $\alpha$ indicates scientist $i$'s $\alpha$-th paper. $\mu_{p}$ is the mean value of all papers' potential impact.
   \item H-index.\\
A scholar has an index value of $H$ if the scholar has $H$ papers with at least $H$ citations. H-index can give an estimate of the impact of a scholar's cumulative research contributions. Based on each selected top conference, we calculate each author's H-index value according to his or her published information in all conferences, then calculate the features of a given institution via the same way as authors' AIF.
\end{itemize}
\textbf{\emph{Geographic distance-based features}}.\\
Given $I$ represents a set of institutions, $I=\{\ I_{1}, I_{2}, \cdots, I_{a}\cdots\}$. $L$ represents a set of the conference location $L=\{\ L_{1}, L_{2}, \cdots, L_{i}\cdots\}$.
For the geographic distance between an institution $I_{a}$ with the conference location $L_{i}$, $d$, can be approximated by a spherical model:
\begin{equation}\label{eq:e2}
    d = 2R \cdot arcsin \sqrt{ sin^2({\frac{\Delta\phi}{2}}) +
        cos(\phi_{a}) \cdot cos(\phi_{i}) \cdot sin^2({\frac{\Delta \lambda}{2}}})
\end{equation}
where $R$ is the Earth's radius, $\phi_{a}$ and $\lambda_{a}$ are the latitude and longitude values of institute $a$, and $\phi_{i}$ and $\lambda_{i}$ are the latitude and longitude values of conference $i$'s location. $\Delta \phi = \left| \phi_{a} - \phi_{i} \right|$ and $\Delta \lambda = \left| \lambda_{a} - \lambda_{i} \right|$.

\begin{figure*}[htbp]
  \centering
  \subfloat[2011]{
    \label{fig:1-a}
    \includegraphics[width=0.23\textwidth]{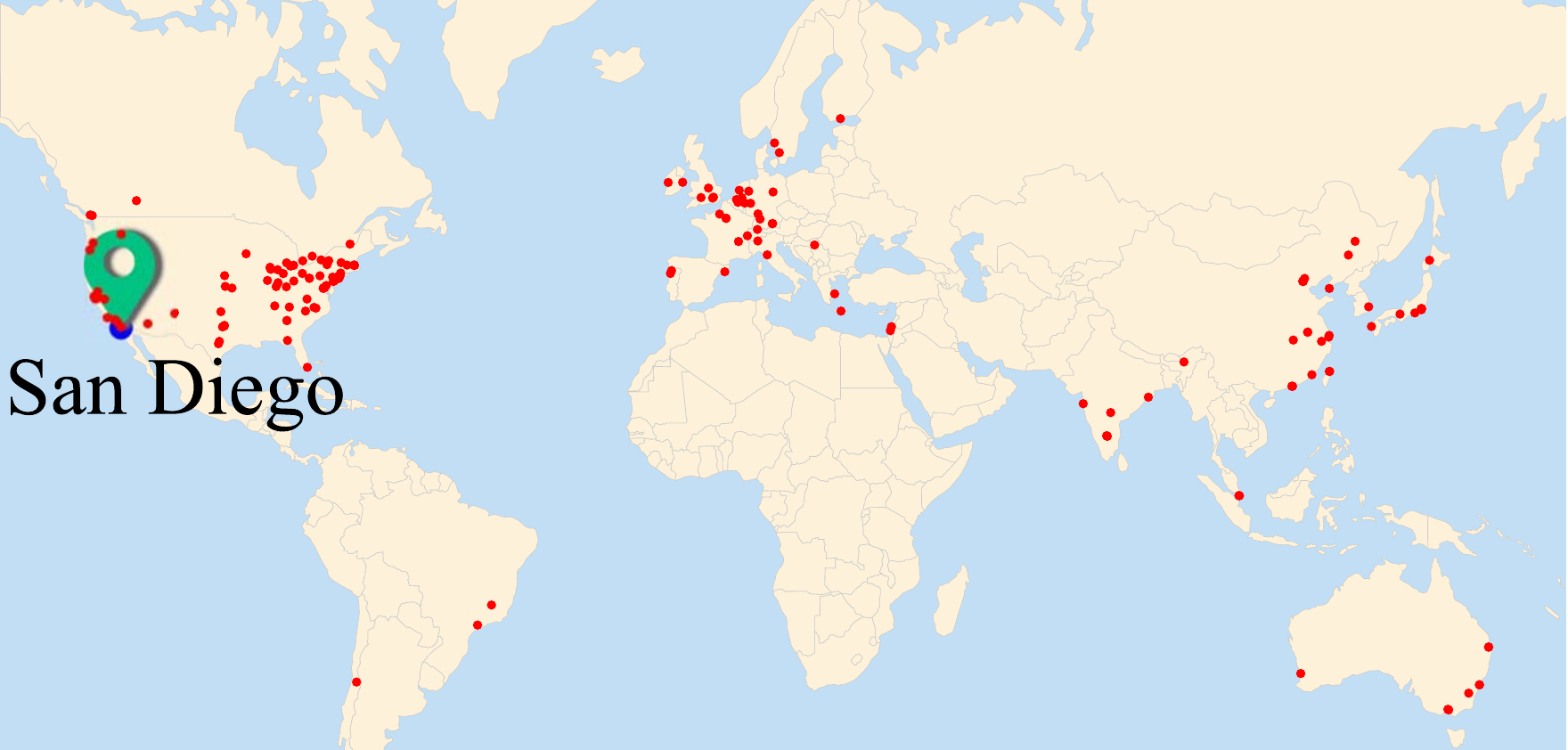}}
  \subfloat[2012]{
    \label{fig:1-b}
    \includegraphics[width=0.23\textwidth]{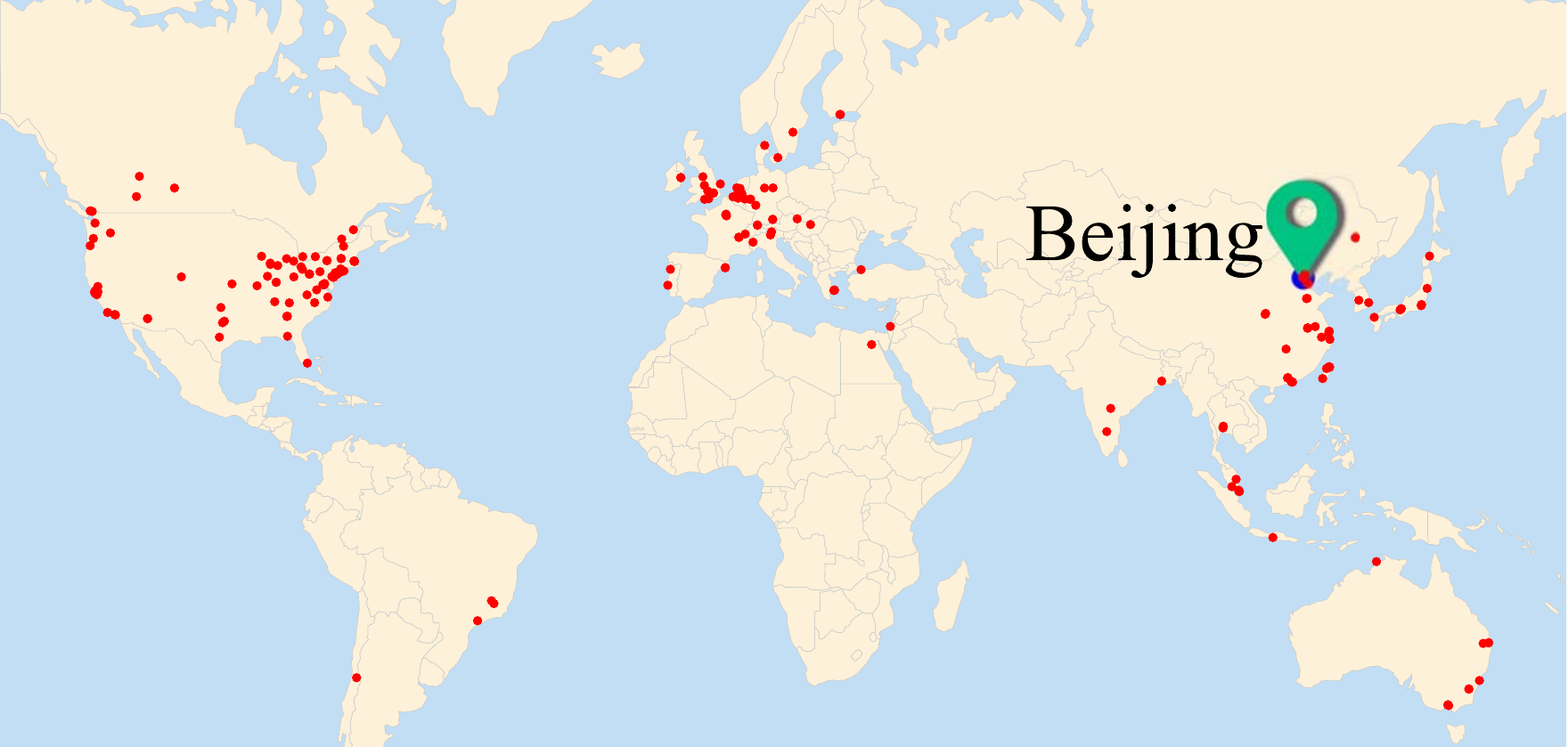}}
  \subfloat[2013]{
    \label{fig:1-c}
    \includegraphics[width=0.23\textwidth]{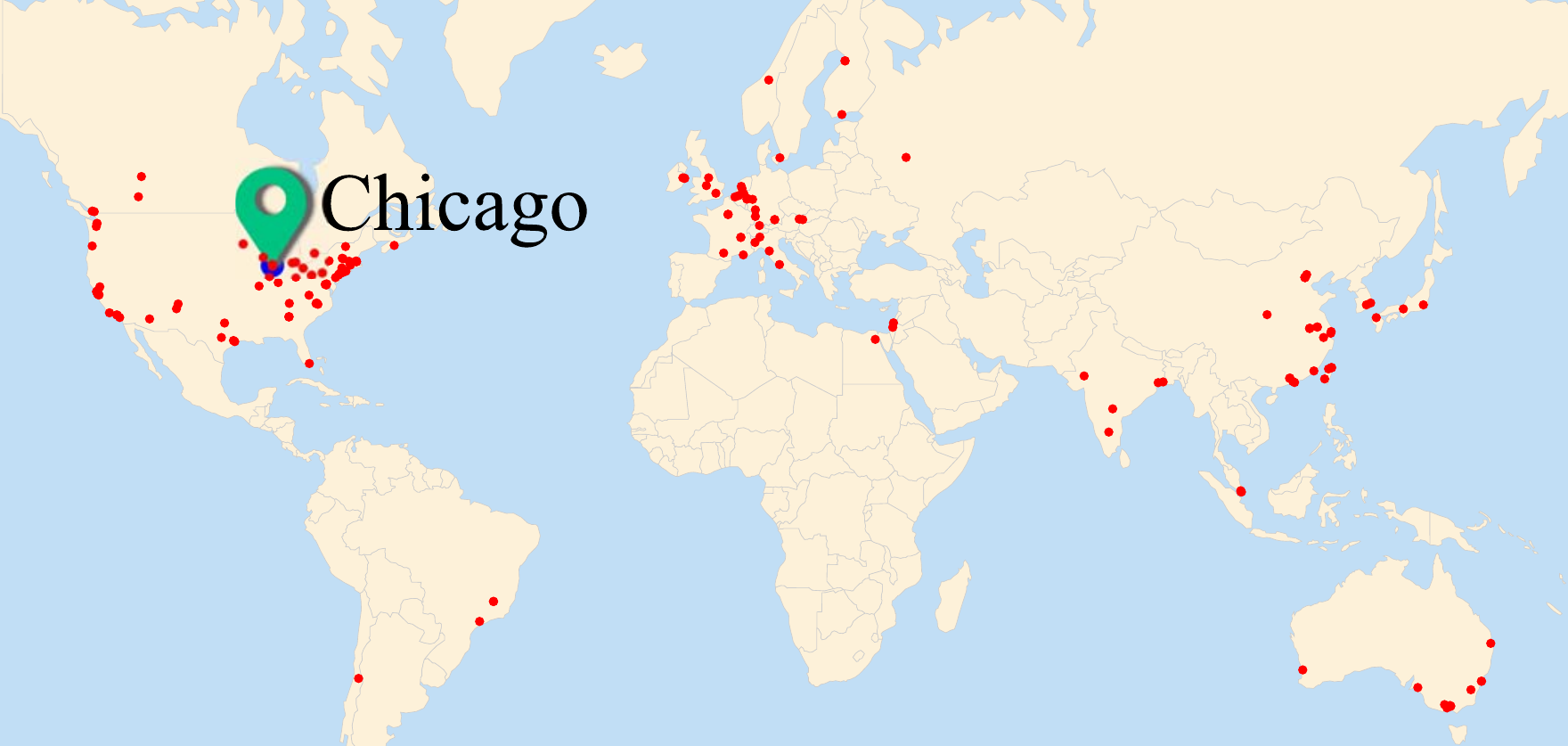}}
  \subfloat[2014]{
    \label{fig:1-d}
    \includegraphics[width=0.23\textwidth]{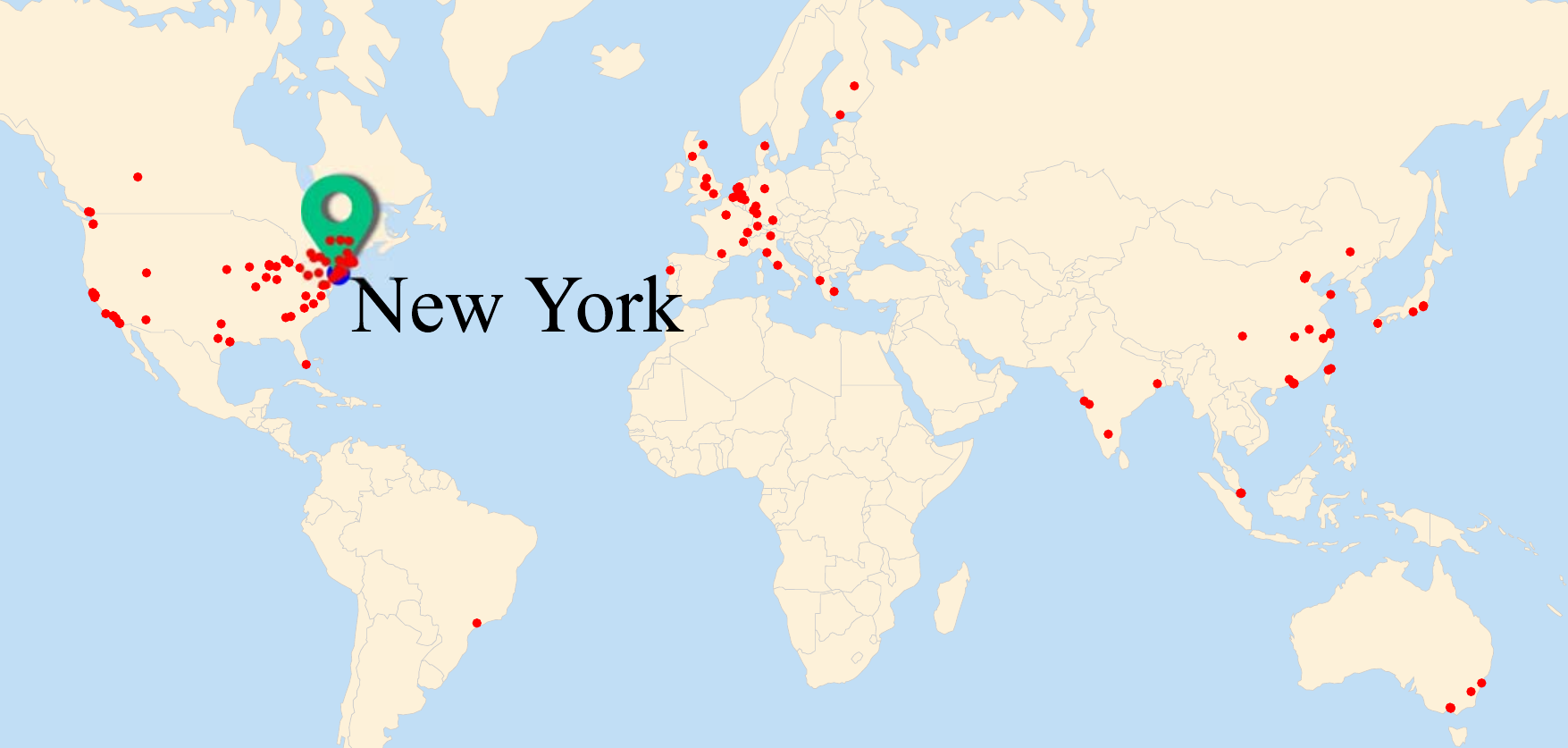}}
  \caption{The coordinate of different institutions published papers on KDD conference from 2011 to 2014.}
  \label{Figure1}
\end{figure*}

\begin{figure}[htbp]
  \centering
  \includegraphics[width=0.45\textwidth]{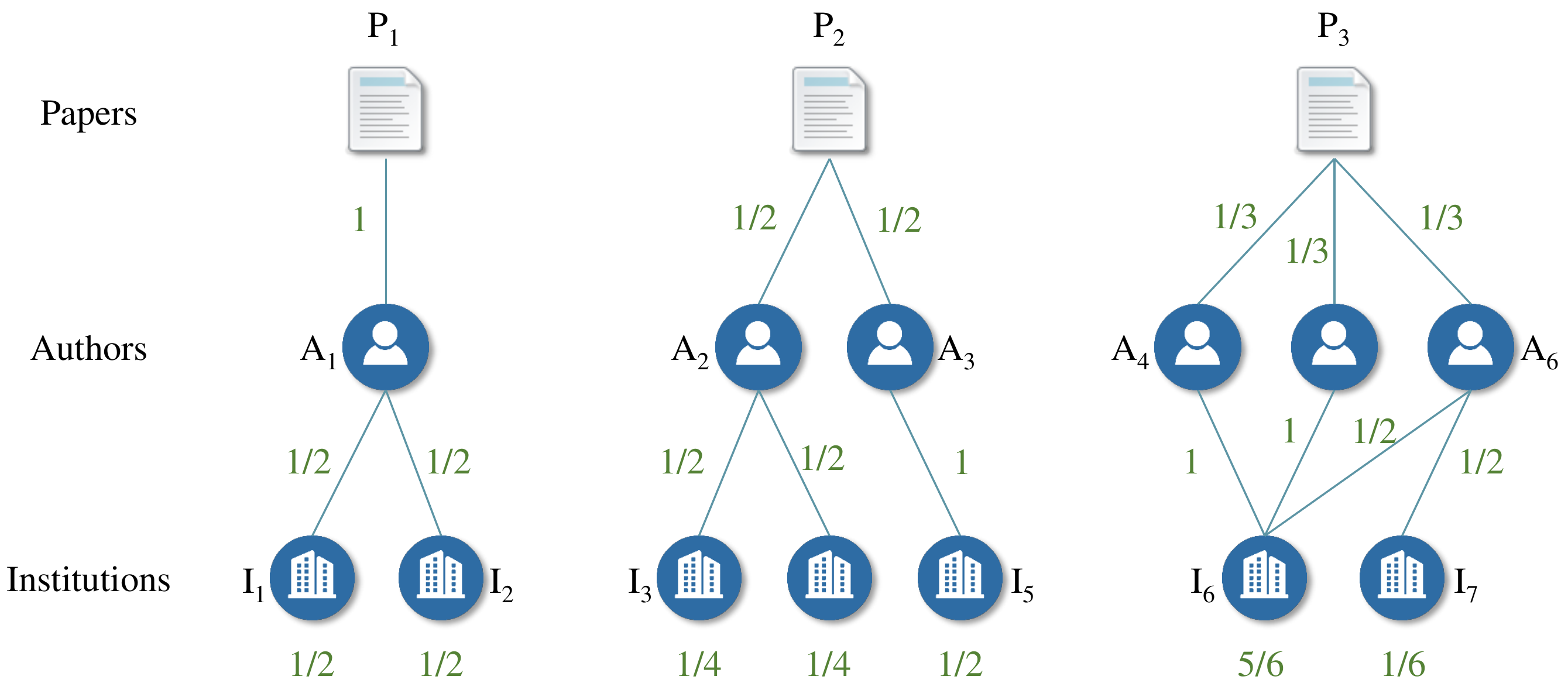}
  \caption{A example demonstrating the relevance scores of each institution.}
  \label{Figure2}
\end{figure}

Figure~\ref{Figure1} shows the geographic distributions of institutions and the location of KDD conference from 2011 to 2014. The red dots represent the institutions, and the green dots represent the conference location. We observe that the authors' institutions published papers on KDD conference mainly distribute in North America, Europe and Asian.
\\
\textbf{\emph{Economic features}}.\\
The GDP per capita of different countries are obtained from the website: http://data.worldbank.org. In our experiments, we use the GDP per capita data from 2000 to 2015.\\ 
\textbf{\emph{Relevance scores-based features}}.\\
Figure~\ref{Figure2} shows an example that the impact of scholarly papers is allocated to different institutions for a given top conference.
The relevance score of each institution in different years can be summarized as follows: (1) each accepted paper is considered as the equal importance. (2) each author has same contribution to a paper. (3) if an author has multiple institutions, each institution also contributes equally.
In addition, we use time trend-based features to weight the relevance scores of each institution in given years. The higher weights are given for the recent years relative to the forecasting year, and the lower weights are given for the years away from predicting year. The weights are normalized, and the sum of weights are set as 1.
The time trend-based features are used for windows of four years. We also use distance trend-based features to weight the relevance scores of each institution in given years. The higher weights are given for the institutions farther from the conference's location in the forecasting year.
\subsection{Feature selection}
In practice, features extracted for machine learning may sometimes be irrelevant or redundant. Selecting the truly correlative features can simplify the predictive model, and even improve the prediction accuracy if there exist some wanted features. To remove the unwanted features, it is necessary to give an importance score for the features of each dimension. In this paper, we use XGBoost feature selection, which can give the score of each feature in the training model, indicating the importance of each feature. The feature score of each feature is the number of times of each feature used in the decision tree nodes partition. The prediction features of impact of institutions are described in Table~\ref{tab:1}.
\begin{table*}[htbp]
  \renewcommand{\arraystretch}{1.2}
  \centering
  \caption{\bf Features used in prediction model.}
  \begin{threeparttable}
    \begin{tabular}{|c|c|c|c|c|c|c|c|}  \hline
      Feature      & Description                           & Feature     & Description\\ \hline
      rel y1       & relevance scores\tnote{1} in y1 year\tnote{2}           & GDP y2      & GDP in y2 year \\ \hline
      rel y2       & relevance scores in y2 year\tnote{2}            & GDP y3      & GDP in y3 year \\ \hline
      rel y3       & relevance scores in y3 year\tnote{2}            & GDP y4      & GDP in y4 year\\ \hline
      rel y4       & relevance scores in y4 year\tnote{2}            & sum(Q)      & sum of Q value\\ \hline
      sum(rel)     & sum of relevance scores                & max(Q)      & maximum of Q value    \\ \hline
      max(rel)     & maximum of relevance scores            & min(Q)      & minimum of Q value \\ \hline
      min(rel)     & minimum of relevance scores            & avg(Q)      & average of Q value\\ \hline
      avg(rel)     & average of relevance scores            & med(Q)      & median of Q value\\ \hline
      med(rel)     & median of relevance scores             & dev(Q)      & deviation of Q value\\ \hline
      dev(rel)     & deviation of relevance scores          & sum(AIF)    & sum of AIF \\ \hline
      wt(rel)      & time weighted relevance scores        & max(AIF)    & maximum of AIF \\ \hline
      wd(rel)      & distance weighted relevance scores     & min(AIF)    & minimum of AIF  \\ \hline
      sum(H-index) & sum of H-index                        & avg(AIF)    & average of AIF\\ \hline
      max(H-index) & maximum of H-index                    & med(AIF)    & median of AIF \\ \hline
      min(H-index) & minimum of H-index                    & dev(AIF)    & deviation of AIF\\ \hline
      avg(H-index) & average of H-index                    & distance y1 & distance between conference and institution in y1 year\\ \hline
      med(H-index) & median of H-index                     & distance y2 & distance between conference and institution in y2 year \\ \hline
      dev(H-index) & deviation of H-index                  & distance y3 & distance between conference and institution in y3 year \\ \hline
      GDP y1       & GDP in y1 year                        & distance y4 & distance between conference and institution in y4 year\\ \hline
    \end{tabular}
    \begin{tablenotes}
      \footnotesize
      \item[1] relevance scores are the number of the accepted papers for each institution in given year.
      \item[2] y year represents the predictive year, y1=y-4, y2=y-3, y3=y-2, y4=y-1.
    \end{tablenotes}
  \end{threeparttable}
  \label{tab:1}
\end{table*}

\subsection{Learning algorithms}
In this section, we describe two algorithms for learning and predicting the impact of institutions for top conferences. One is Gradient boosting decision trees (GBDT), which is a comparison model. The other is XGBoost, which is intended to enhance the prediction performance for predicting the impact of institutions in the future.
\subsubsection{GBDT}
The GBDT model is used to predict the number of accepted papers for each institution in the eight top conferences in next year. GBDT model suits for dealing with a mass of features and no-linear relationships between the predictor variables and the target variable. GBDT model is the extension of weak decision trees, and the error function selected is the mean square error function when GBDT is used to regression problem. In dealing with the problem of predicting the impact of institutions in next year, Sandulescu et al.~\cite{sandulescu2016predicting} used three kinds of features: statistics-based features, trend-based features and AIF-based features including time weighted relevance scores and the sum, maximum, minimum, median, average, deviation of relevance scores and authors' AIF.
\subsubsection{XGBoost}
XGBoost is a scalable end-to-end tree boosting system, which runs more than ten times faster than existing currently popular solutions.

XGBoost's tree boosting mainly includes two parts. One is regularized learning objective, the other is gradient tree boosting. For a given dataset $D$ with $m$ examples and $n$ features D=\{$(x_{i},y_{i})$\}, where $|D|=m, x_{i} \in R^{n}, y_{i} \in R$, 
the formula of output prediction is as follows:
\begin{equation}\label{eq:e3}
  \hat{y_{i}}=\phi(x_{i})=\sum_{k=1}^{K}f_{k}(x_{i}), f_{k}\in F,
\end{equation}
where $F=\{f_{x}=\omega_{q(x)}\}(q:R^{n}\rightarrow T, \omega\in R^{T})$ is the space of regression trees. $q$ indicates the structure of each tree, which maps an example to the corresponding leaf index. $T$ is the number of leaves of the tree. $f_{k}$ is consistent with tree structure $q$ and leaf weights $\omega$. In order to learn $f_{k}$, the following regularized objective is introduced:
\begin{equation}\label{eq:e4}
  \Gamma (\phi)=\sum_{i}l\left (\hat{y_{i}},y_{i} \right )+\sum_{k}\Omega \left ( f_{k} \right ),
\end{equation}
where $\Omega \left (f \right )=\gamma{T}+\frac{1}{2}\lambda\|\omega\|^{2}$,
$\Omega$ controls the complexity of the model to avoid over-fitting. $l$ indicates a loss function, which calculates the difference between the prediction value $\hat{y_{i}}$ and the true value $y_{i}$.

To improve the prediction model according to the equation, $f_{t}$ is added: 
\begin{equation}\label{eq:e5}
  \Gamma ^{\left ( t \right )}=\sum_{i=1}^{n}l(y_{i},\hat{y_{i}}^{(t-1)}+f_{t}(x_{i}))+\Omega(f_{t}),
\end{equation}
where $\hat{y_{i}}^{(t)}$ is the prediction of the $i^{th}$ instance at the $t^{th}$ iteration.

In the XGBoost model, the second order Taylor expansion is used to accelerate the optimization procedure of the objective function, which is shown as follows:
\begin{equation}\label{eq:e6}
  \widetilde{\Gamma}^{(t)}=\sum_{i=1}^{n}[g_{i}f_{t}(x_{i})+\frac{1}{2}h_{i}f_{t}^{2}(x_{i})]+\gamma T+\frac{1}{2}\lambda\sum_{j=1}^{T}w_{j}^{2},
\end{equation}
where $g_{i}=\partial_{\hat{y}(t-1)}l(y_{i},\hat{y_{i}}^{(t-1)})$, and $h_{i}=\partial_{\hat{y}(t-1)}^{2}l(y_{i},\hat{y_{i}}^{(t-1)})$ are the first and second order gradient statistics for the loss function.

For an independent tree structure $q(x)$, the optimal weight $w_{j}^{\ast}$ of leaf $j$ is calculated as follows:
\begin{equation}\label{eq:e7}
  \omega _{j}^{\ast }=-\frac{\sum_{i\in I_{j}}g_{i}}{\sum_{i\in I_{j}}h_{i}+\lambda }.
\end{equation}

The corresponding optimal objective function is calculated by the following formula:
\begin{equation}\label{eq:e8}
  \widetilde{\Gamma}^{\left ( t \right )} =-\frac{1}{2}\sum_{j=1}^{T}\frac{(\sum_{i\in I_{j}}g_{i})^{2}}{\sum_{i\in I_{j}}h_{i}+\lambda}+\gamma T.
\end{equation}

Our training data is composed of a sequence of relevant features of the institutional impact $\Lambda=(\theta_{1}, \theta_{2}, \theta_{3}, \cdots, \theta_{n})$, where $\theta_{i}$ represents the $i$-th feature. In our experiments, we mainly use historical data in the previous four year to predict the number of the accepted papers in the next year from each institution, and the experiments are conducted across years 2000 to 2010. The features include author-based features, relevance scores-base features, geographic distance-based features, and GDP-based features. We use the number of accepted papers of each institution in prediction year to label each training data to train XGBoost model for the task of institution impact prediction. Our testing data is extracted from the selected conference between 2011 and 2015. We use the extracted features from 2011 to 2014 to predict the number of the accepted papers in 2015. We employ the XGBoost algorithm that can yield accurate prediction for the future impact of each institution.

\subsection{Normalized Discounted Cumulative Gain (NDCG) for model evaluation}
NDCG is a normalized measure method of Discounted Cumulative Gain (DCG). NDCG usually is used to indicate the accuracy of prediction model. DCG is given by:
\begin{equation}\label{eq:e9}
  DCG_{n}=\sum_{i=1}^{n}\frac{rel_{i}}{log_{2}^{i+1}},
\end{equation}
where $DCG_{n}$ is the weighted sum of relevant degree of ranked entities, and its weight is a decreasing function varying according to the ranked position. Variable $i$ is the ranking of an institution, and $rel_{i}$ is the relevance score of the $i$-th ranked institution.
Via normalizing DCG values, NDCG{@}N is given by:
\begin{equation}\label{eq:e10}
  NDCG_{n}=\frac{DCG_{n}}{IDCG_{n}},
\end{equation}
where $IDCG$ is an ideal DCG, which is considered as the simple DCG measure with the best ranking results. Therefore, the probability score of NDCG measurement always are between $0$ and $1$. In this paper, NDCG reflects the importance of an institution in the given relevant top conference. If an institution is not appeared in the final ranking results, its NDCG value will be set as $0$.

\section{Results}
\begin{figure*}[!ht]
  \centering
  \includegraphics[width=\textwidth]{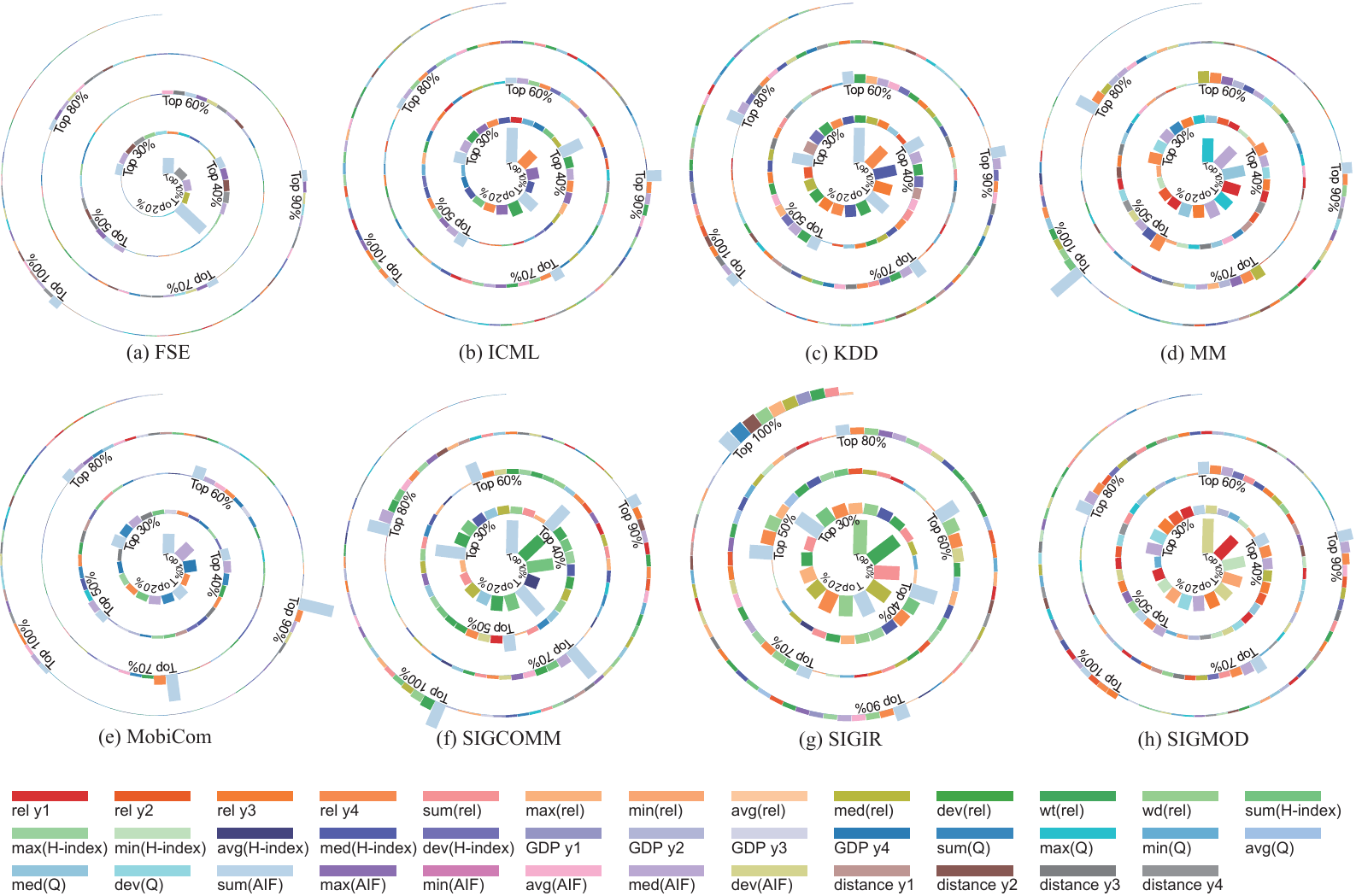}
  \caption{Feature importance ranking from four features for different conferences.\\ Notes: the horizontal axis indicates the relevant importance scores, and the vertical axis indicates the features.}
  \label{Figure3}
\end{figure*}

Considering the historical relevance scores of the accepted papers is a common practice in predicting institutional contribution in a conference~\cite{sandulescu2016predicting}.
Via feature selection, we find that author-based features introduce more impact, and we observe that the authors' impact of each institution for some top conferences dominates the impact of the institution. In our experiments, we train the predictive model using top $10\%$, top $20\%$, top $30\%$,$\cdots$, and all features including author-based features, relevance scores-based features, geographic distance-based features, and state GDP-based features. Via feature selection, feature importance ranking of different features for the selected conferences is shown in Figure~\ref{Figure3}. The feature importance ranking indicates the best predictive performance according to the selected features. In Figure~\ref{Figure3}, we observe that the feature importance scores are different according to features selected to train the predictive model.

\begin{table*}[htbp]
  \scriptsize
  \centering
  \caption{\bf Four features' importance scores.}
  \subfloat[FSE]{
    \begin{tabular}{|p{0.75cm}<{\centering}|p{0.65cm}<{\centering}|p{0.9cm}<{\centering}|p{0.8cm}<{\centering}|p{0.6cm}<{\centering}|}
    \hline
      Features&Author&Relevance&Distance&GDP   \\
              &      &scores  &        &      \\ \hline
      10\%    &0.6349&0.1451  &0.2200  &0.0000\\ \hline
      20\%    &0.9944&0.0029  &0.0026  &0.0000\\ \hline
      30\%    &0.5738&0.1999  &0.2262  &0.0000\\ \hline
      40\%    &0.5926&0.0948  &0.3127  &0.0000\\ \hline
      50\%    &0.5660&0.1149  &0.2353  &0.0837\\ \hline
      60\%    &0.5679&0.0913  &0.2591  &0.0819\\ \hline
      70\%    &0.6087&0.2068  &0.1325  &0.0520\\ \hline
      80\%    &0.5679&0.1953  &0.1572  &0.0796\\ \hline
      90\%    &0.5890&0.2307  &0.1222  &0.0583\\ \hline
      100\%   &0.5616&0.1765  &0.2125  &0.0493\\ \hline
    \end{tabular}
    \label{tab:2}
  }
  \subfloat[ICML]{
    \begin{tabular}{|p{0.75cm}<{\centering}|p{0.65cm}<{\centering}|p{0.9cm}<{\centering}|p{0.8cm}<{\centering}|p{0.6cm}<{\centering}|} \hline
      Features&Author&Relevance&Distance&GDP   \\
              &      &scores  &        &      \\ \hline
      10\%    &0.7547&0.2452  &0.0000  &0.0000\\ \hline
      20\%    &0.6272&0.3041  &0.0000  &0.0687\\ \hline
      30\%    &0.5966&0.3299  &0.0000  &0.0735\\ \hline
      40\%    &0.6198&0.3695  &0.0000  &0.0107\\ \hline
      50\%    &0.6174&0.3395  &0.0000  &0.0431\\ \hline
      60\%    &0.5553&0.3805  &0.0379  &0.0263\\ \hline
      70\%    &0.5464&0.3822  &0.0243  &0.0470\\ \hline
      80\%    &0.5266&0.3790  &0.0584  &0.0363\\ \hline
      90\%    &0.5814&0.3049  &0.0726  &0.0412\\ \hline
      100\%   &0.4883&0.3676  &0.0976  &0.0465\\ \hline
    \end{tabular}
    \label{tab:3}
  }
  \subfloat[KDD]{
    \begin{tabular}{|p{0.75cm}<{\centering}|p{0.65cm}<{\centering}|p{0.9cm}<{\centering}|p{0.8cm}<{\centering}|p{0.6cm}<{\centering}|} \hline
      Features&Author&Relevance&Distance&GDP   \\
              &      &scores  &        &      \\ \hline
      10\%    &0.5839&0.4160  &0.0000  &0.0000\\ \hline
      20\%    &0.3518&0.6483  &0.0000  &0.0000\\ \hline
      30\%    &0.4491&0.4352  &0.1155  &0.0000\\ \hline
      40\%    &0.5438&0.3770  &0.0792  &0.0000\\ \hline
      50\%    &0.4746&0.4155  &0.1097  &0.0000\\ \hline
      60\%    &0.4686&0.4163  &0.0861  &0.0289\\ \hline
      70\%    &0.4994&0.3182  &0.1446  &0.0379\\ \hline
      80\%    &0.5378&0.2950  &0.1324  &0.0347\\ \hline
      90\%    &0.5655&0.2764  &0.1252  &0.0328\\ \hline
      100\%   &0.4663&0.3460  &0.1535  &0.0343\\ \hline
    \end{tabular}
    \label{tab:4}
  }
  \\
  \subfloat[MM]{
     \begin{tabular}{|p{0.75cm}<{\centering}|p{0.65cm}<{\centering}|p{0.9cm}<{\centering}|p{0.8cm}<{\centering}|p{0.6cm}<{\centering}|} \hline
      Features&Author&Relevance&Distance&GDP   \\
              &      &scores  &        &      \\ \hline
      10\%    &0.7986&0.2014  &0.0000  &0.0000\\ \hline
      20\%    &0.5819&0.4182  &0.0000  &0.0000\\ \hline
      30\%    &0.5751&0.4250  &0.0000  &0.0000\\ \hline
      40\%    &0.4351&0.3764  &0.1884  &0.0000\\ \hline
      50\%    &0.4925&0.3213  &0.1863  &0.0000\\ \hline
      60\%    &0.4716&0.3697  &0.0865  &0.0721\\ \hline
      70\%    &0.4295&0.4025  &0.0833  &0.0848\\ \hline
      80\%    &0.5075&0.2867  &0.0739  &0.1322\\ \hline
      90\%    &0.4846&0.3326  &0.1166  &0.0662\\ \hline
      100\%   &0.6690&0.3285  &0.0017  &0.0005\\ \hline
    \end{tabular}
    \label{tab:5}
  }
  \subfloat[MobiCom]{
     \begin{tabular}{|p{0.75cm}<{\centering}|p{0.65cm}<{\centering}|p{0.9cm}<{\centering}|p{0.8cm}<{\centering}|p{0.6cm}<{\centering}|} \hline
      Features&Author&Relevance&Distance&GDP   \\
              &      &scores  &        &      \\ \hline
      10\%    &0.6609&0.1069  &0.0000  &0.2323\\ \hline
      20\%    &0.7424&0.1079  &0.0675  &0.0821\\ \hline
      30\%    &0.7267&0.0688  &0.0865  &0.1180\\ \hline
      40\%    &0.6136&0.1382  &0.1241  &0.1240\\ \hline
      50\%    &0.6360&0.1704  &0.1071  &0.0863\\ \hline
      60\%    &0.5767&0.1855  &0.0520  &0.1857\\ \hline
      70\%    &0.6503&0.2559  &0.0212  &0.0424\\ \hline
      80\%    &0.5932&0.2067  &0.1028  &0.0973\\ \hline
      90\%    &0.7889&0.1165  &0.0680  &0.0265\\ \hline
      100\%   &0.5201&0.2693  &0.1414  &0.0691\\ \hline
    \end{tabular}
    \label{tab:6}
  }
  \subfloat[SIGCOMM]{
     \begin{tabular}{|p{0.75cm}<{\centering}|p{0.65cm}<{\centering}|p{0.9cm}<{\centering}|p{0.8cm}<{\centering}|p{0.6cm}<{\centering}|} \hline
      Features&Author&Relevance&Distance&GDP   \\
              &      &scores  &        &      \\ \hline
      10\%    &0.7282&0.2718  &0.0000  &0.0000\\ \hline
      20\%    &0.6050&0.3949  &0.0000  &0.0000\\ \hline
      30\%    &0.6077&0.3924  &0.0000  &0.0000\\ \hline
      40\%    &0.6347&0.3653  &0.0000  &0.0000\\ \hline
      50\%    &0.5527&0.4474  &0.0000  &0.0000\\ \hline
      60\%    &0.5559&0.4442  &0.0000  &0.0000\\ \hline
      70\%    &0.7734&0.2264  &0.0000  &0.0000\\ \hline
      80\%    &0.6440&0.2636  &0.0922  &0.0000\\ \hline
      90\%    &0.4482&0.3720  &0.1291  &0.0508\\ \hline
      100\%   &0.5137&0.4024  &0.0354  &0.0484\\ \hline
    \end{tabular}
    \label{tab:7}
  }
  \\
  \subfloat[SIGIR]{
     \begin{tabular}{|p{0.75cm}<{\centering}|p{0.65cm}<{\centering}|p{0.9cm}<{\centering}|p{0.8cm}<{\centering}|p{0.6cm}<{\centering}|} \hline
      Features&Author&Relevance&Distance&GDP   \\
              &      &scores  &        &      \\ \hline
      10\%    &0.0000&1.0000  &0.0000  &0.0000\\ \hline
      20\%    &0.2068&0.7932  &0.0000  &0.0000\\ \hline
      30\%    &0.4932&0.5067  &0.0000  &0.0000\\ \hline
      40\%    &0.5681&0.4320  &0.0000  &0.0000\\ \hline
      50\%    &0.4802&0.5197  &0.0000  &0.0000\\ \hline
      60\%    &0.5035&0.4964  &0.0000  &0.0000\\ \hline
      70\%    &0.5437&0.4253  &0.0311  &0.0000\\ \hline
      80\%    &0.4988&0.3845  &0.1168  &0.0000\\ \hline
      90\%    &0.4764&0.3616  &0.0897  &0.0725\\ \hline
      100\%   &0.2928&0.5646  &0.0717  &0.0710\\ \hline
    \end{tabular}
    \label{tab:8}
  }
  \subfloat[SIGMOD]{
     \begin{tabular}{|p{0.75cm}<{\centering}|p{0.65cm}<{\centering}|p{0.9cm}<{\centering}|p{0.8cm}<{\centering}|p{0.6cm}<{\centering}|} \hline
      Features&Author&Relevance&Distance&GDP   \\
              &      &scores  &        &      \\ \hline
      10\%    &0.5624&0.4377  &0.0000  &0.0000\\ \hline
      20\%    &0.6184&0.3816  &0.0000  &0.0000\\ \hline
      30\%    &0.5646&0.3765  &0.0000  &0.0591\\ \hline
      40\%    &0.4732&0.4470  &0.0400  &0.0399\\ \hline
      50\%    &0.4826&0.4269  &0.0552  &0.0351\\ \hline
      60\%    &0.5033&0.3802  &0.0889  &0.0277\\ \hline
      70\%    &0.4840&0.3575  &0.1200  &0.0385\\ \hline
      80\%    &0.4774&0.3762  &0.1088  &0.0375\\ \hline
      90\%    &0.4915&0.3655  &0.1062  &0.0368\\ \hline
      100\%   &0.4560&0.3753  &0.1323  &0.0362\\ \hline
    \end{tabular}
    \label{tab:9}
  }
\end{table*}

Figure~\ref{Figure3}a shows the feature importance ranking via different percentages of features selection for FSE. We observe that the features related to authors' AIF rank first for using different percentages of features to train the predictive model. Author-based features' importance scores, relevance scores-based features' importance scores, geographical distance-based features' importance scores and GDP-based features' importance scores for FSE are listed in Table~\ref{tab:2}. In Table~\ref{tab:2}, we observe that author-based features' importance scores are the highest, which are higher than 0.5. Specially, author-based features' importance scores reach 0.9944 via using top $20\%$ features to train the predictive model. Relevance scores-based features' importance scores are lower than 0.25, and geographical distance-based features' importance scores are lower than 0.32. The lowest feature importance scores are from GDP-based features, which are lower than 0.1.

Figure~\ref{Figure3}b shows feature importance ranking for ICML. We observe that the sum of authors' AIF features rank first in all the features' importance ranking list for ICML. In Table~\ref{tab:3}, we observe that author-based features' importance scores are the highest in the four kinds of features: author-based features, relevance scores-based features, geographical distance-based features and GDP-based features. All the author-based features' importance scores exceed 0.48 for using different percentages of features to train the predictive model. Relevance scores-based features' importance scores are between 0.2 and 0.4. For ICML, geographical distance-based features begin to take effect from top $60\%$ features, and GDP-based features' importance scores are from top $20\%$ features.

Figure~\ref{Figure3}c illustrates the feature importance ranking for KDD. We observe that the sum of authors' AIF features ranks first in terms of feature importance. From Table~\ref{tab:4}, we observe that author-based features' importance scores are higher than relevance scores-based features' importance scores except for the top $20\%$ features. Author-based features' importance scores range from 0.3 to 0.6, and relevance scores-based features' importance scores are between 0.2 and 0.7. Compared to ICML, geographical distance-based features begin to take effect earlier when the top $30\%$ features are used. GDP-based features are found after choosing the top $60\%$ features.

Figure~\ref{Figure3}d shows the feature importance ranking for MM. Compared with FSE, ICML and KDD, the relevant features with authors do not always appear first on the feature importance ranking list, other relevant features such as the historical scores of institutions rank first for top $40\%$, top $50\%$, top $60\%$ and top $70\%$ features. In Table~\ref{tab:5}, we observe that author-based features' importance scores are between 0.4 and 0.8, and relevance scores-based features' importance scores are between 0.2 and 0.5.

Figure~\ref{Figure3}e illustrates the feature importance ranking for MobiCom. According to Figure~\ref{Figure3}e, we observe that authors' AIF features always rank first in feature importance ranking list. In Table~\ref{tab:6}, we observe that author-based features' importance scores are higher than the relevance scores-based features' importance scores. Like FSE, author-based features' importance scores are higher than 0.5 for using different percentages of features to train the model. The relevance scores-based features' importance scores are between 0.1 and 0.3. The geographical distance-based features' importance scores and GDP-based features' importance scores range from 0 to 0.3. Compared to FSE, ICML, KDD, and MM, GDP-based features' importance scores begin to take effect from using top $10\%$ features to train the model.

Figure~\ref{Figure3}f illustrates the feature importance ranking for SIGCOMM. According to Figure~\ref{Figure3}f, we observe that authors' AIF features rank first in feature importance ranking list. Like MobiCom, authors-based features' importance scores are higher than the relevance scores-based features' importance scores for SIGCOMM, which is shown in Table~\ref{tab:7}. In Table~\ref{tab:7}, we observe that geographical distance-based features and GDP-based features have quite a few effects for improving the predictive model.

Figure~\ref{Figure3}g shows the feature importance ranking for SIGIR. According to Figure~\ref{Figure3}g, we observe that the authors-based features rank first excluding the top $10\%$ features. In Table~\ref{tab:8}, we observe that author-based features' importance scores are between 0 and 0.6, and the relevance score-based features' importance scores are between 0.3 and 1. For less than top $30\%$ features, the relevant score-based features' importance scores are higher than authors-based features' importance scores. Like SIGCOMM, geographical distance-based features and GDP-based features have quite a few effects for improving the performance of predictive model.

Figure~\ref{Figure3}h shows the feature importance ranking for SIGMOD. According to Figure~\ref{Figure3}h, we observe that authors' AIF features rank first in feature importance ranking list excluding the case of using all features to train predictive model. In Table~\ref{tab:9}, we observe that author-based features' importance scores are higher than relevance scores-based features' importance scores. Author-based features' importance scores are between 0.4 and 0.7, and the relevance scores-based features' importance scores are between 0.3 and 0.5.


Because we find that the historical relevance scores of each institution are not the most important factors for predicting the number of the accepted papers via the feature selection, this drives us to explore the performance of the prediction model without considering the relevant features with the historical relevance scores of each institution. We resume to train the model using features excluding the features relevant to the historical relevance scores of each institution.

\begin{figure*}[!ht]
  \centering
  \includegraphics[width=\textwidth]{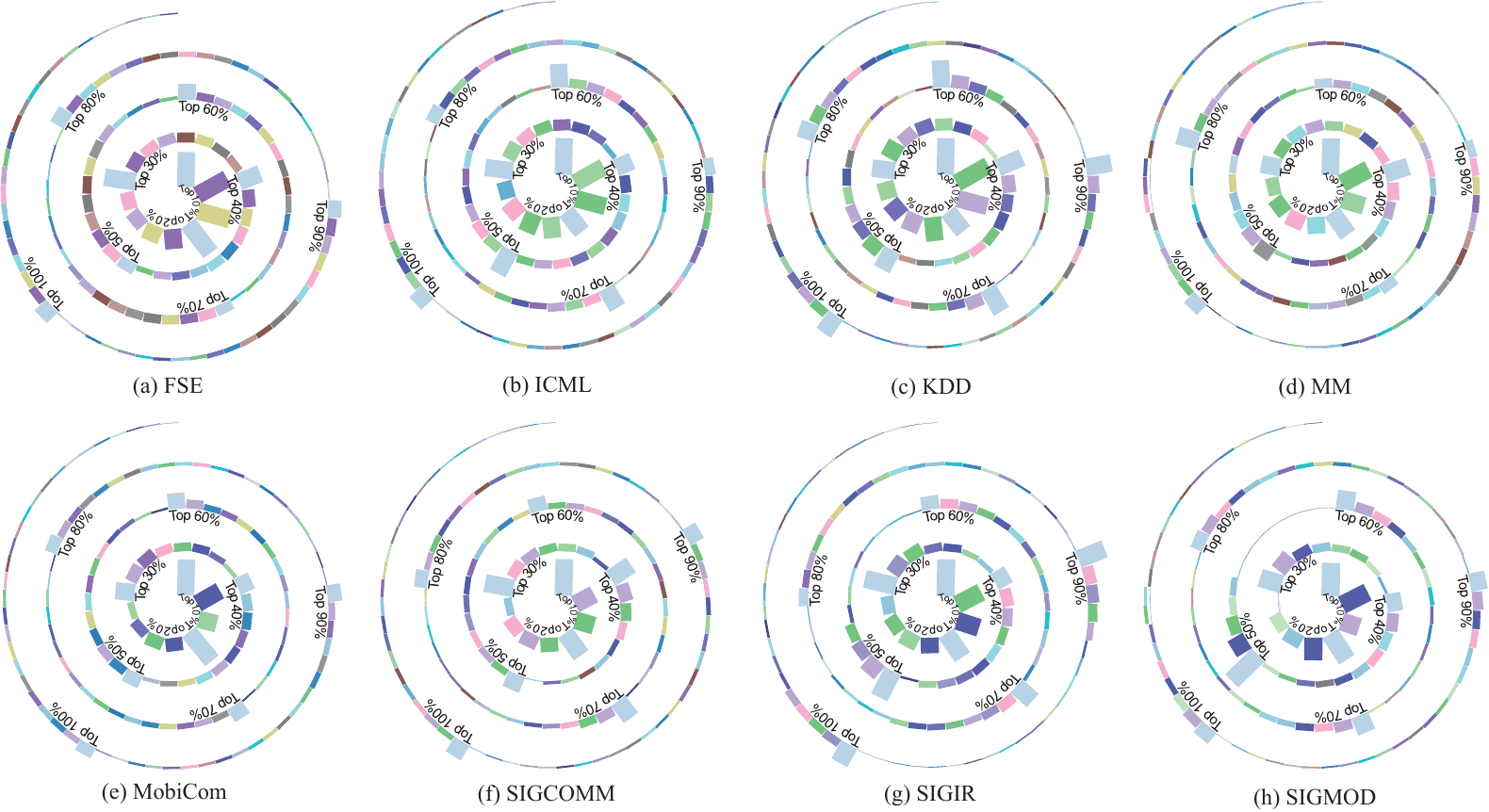}
  \caption{Feature importance ranking from three features for different conferences.\\
  Notes: the horizontal axis indicates the relevant importance scores, and the vertical axis indicates the features.}
  \label{Figure4}
\end{figure*}

Figure~\ref{Figure4} shows the feature importance ranking according to different percentages of features excluding the relevant features with the relevant historical scores of each institution.

Figure~\ref{Figure4}a shows the feature importance ranking for FSE. According to Figure~\ref{Figure4}a, we observe that authors' AIF features rank first for different percentages of features. In Table~\ref{tab:10}, we observe that author-based features' importance scores are the highest, which are between 0.6 and 1. Geographical distance-based features' importance scores are between 0 and 0.4, and GDP-based features' importance scores are between 0 and 0.07.

\begin{table*}[htbp]
  \scriptsize
  \centering
  \caption{\bf Three features' importance scores.}
  \subfloat[FSE]{
    \begin{tabular}{|c|c|c|c|} \hline
      Features&Author&Distance&GDP\\ \hline
      10\%      &1.0000&0.0000&0.0000\\ \hline
      20\%      &1.0000&0.0000&0.0000\\ \hline
      30\%      &0.7411&0.2589&0.0000\\ \hline
      40\%      &1.0000&0.0000&0.0000\\ \hline
      50\%      &0.6434&0.3566&0.0000\\ \hline
      60\%      &0.7337&0.2186&0.0479\\ \hline
      70\%      &0.6187&0.1636&0.0508\\ \hline
      80\%      &0.7300&0.2029&0.0672\\ \hline
      90\%      &0.7102&0.2128&0.0224\\ \hline
      100\%    &0.7868&0.1526&0.0608\\ \hline
    \end{tabular}
    \label{tab:10}
  }
  \subfloat[ICML]{
    \begin{tabular}{|c|c|c|c|} \hline
      Features&Author&Distance&GDP\\ \hline
      10\%      &1.0000&0.0000&0.0000\\ \hline
      20\%      &1.0000&0.0000&0.0000\\ \hline
      30\%      &1.0000&0.0000&0.0000\\ \hline
      40\%      &1.0000&0.0000&0.0000\\ \hline
      50\%      &0.9283&0.0718&0.0000\\ \hline
      60\%      &0.9509&0.0491&0.0000\\ \hline
      70\%      &0.9352&0.0648&0.0000\\ \hline
      80\%      &0.8677&0.1324&0.0000\\ \hline
      90\%      &0.7978&0.1586&0.0335\\ \hline
      100\%    &0.8355&0.1065&0.0580\\ \hline
    \end{tabular}
    \label{tab:11}
  }
  \subfloat[KDD]{
    \begin{tabular}{|c|c|c|c|} \hline
      Features&Author&Distance&GDP\\ \hline
      10\%      &1.0000&0.0000&0.0000\\ \hline
      20\%      &1.0000&0.0000&0.0000\\ \hline
      30\%      &1.0000&0.0000&0.0000\\ \hline
      40\%      &0.8923&0.1076&0.0000\\ \hline
      50\%      &0.8840&0.1160&0.0000\\ \hline
      60\%      &0.8367&0.1632&0.0000\\ \hline
      70\%      &0.8359&0.1641&0.0000\\ \hline
      80\%      &0.8563&0.0933&0.0504\\ \hline
      90\%      &0.8112&0.1342&0.0545\\ \hline
      100\%    &0.8129&0.1284&0.0585\\ \hline
    \end{tabular}
    \label{tab:12}
  }
  \\
  \subfloat[MM]{
    \begin{tabular}{|c|c|c|c|} \hline
      Features&Author&Distance&GDP\\ \hline
      10\%      &1.0000&0.0000&0.0000\\ \hline
      20\%      &1.0000&0.0000&0.0000\\ \hline
      30\%      &1.0000&0.0000&0.0000\\ \hline
      40\%      &0.8536&0.1464&0.0000\\ \hline
      50\%      &0.8369&0.1631&0.0000\\ \hline
      60\%      &0.7673&0.1647&0.0679\\ \hline
      70\%      &0.7945&0.1376&0.0681\\ \hline
      80\%      &0.7820&0.1110&0.1071\\ \hline
      90\%      &0.6761&0.2509&0.0731\\ \hline
      100\%    &0.7549&0.1567&0.0884\\ \hline
    \end{tabular}
    \label{tab:13}
  }
  \subfloat[MobiCom]{
    \begin{tabular}{|c|c|c|c|} \hline
      Features&Author&Distance&GDP\\ \hline
      10\%      &1.0000&0.0000&0.0000\\ \hline
      20\%      &1.0000&0.0000&0.0000\\ \hline
      30\%      &1.0000&0.0000&0.0000\\ \hline
      40\%      &0.8870&0.0655&0.0474\\ \hline
      50\%      &0.9079&0.0000&0.0920\\ \hline
      60\%      &0.8805&0.0000&0.1194\\ \hline
      70\%      &0.8245&0.0761&0.0996\\ \hline
      80\%      &0.7736&0.1195&0.1069\\ \hline
      90\%      &0.7552&0.1418&0.1030\\ \hline
      100\%    &0.7700&0.1342&0.0954\\ \hline
    \end{tabular}
    \label{tab:14}
  }
  \subfloat[SIGCOMM]{
    \begin{tabular}{|c|c|c|c|} \hline
      Features&Author&Distance&GDP\\ \hline
      10\%      &1.0000&0.0000&0.0000\\ \hline
      20\%      &1.0000&0.0000&0.0000\\ \hline
      30\%      &1.0000&0.0000&0.0000\\ \hline
      40\%      &0.9421&0.0579&0.0000\\ \hline
      50\%      &0.9260&0.0000&0.0074\\ \hline
      60\%      &0.8731&00550&0.0719\\ \hline
      70\%      &0.8703&0.0345&0.0954\\ \hline
      80\%      &0.7940&0.0949&0.0671\\ \hline
      90\%      &0.7998&0.1222&0.0779\\ \hline
      100\%    &0.8227&0.1063&0.0709\\ \hline
    \end{tabular}
    \label{tab:15}
  }
  \\
  \subfloat[SIGIR]{
    \begin{tabular}{|c|c|c|c|} \hline
      Features&Author&Distance&GDP\\ \hline
      10\%      &1.0000&0.0000&0.0000\\ \hline
      20\%      &1.0000&0.0000&0.0000\\ \hline
      30\%      &0.8220&0.0000&0.1779\\ \hline
      40\%      &0.9120&0.0000&0.0880\\ \hline
      50\%      &0.8638&0.0000&0.1363\\ \hline
      60\%      &0.9026&0.0000&0.0972\\ \hline
      70\%      &0.8772&0.0000&0.1229\\ \hline
      80\%      &0.8991&0.0000&0.1008\\ \hline
      90\%      &0.8354&0.0196&0.1450\\ \hline
      100\%    &0.7953&0.0541&0.1506\\ \hline
    \end{tabular}
    \label{tab:16}
  }
  \subfloat[SIGMOD]{
    \begin{tabular}{|c|c|c|c|} \hline
      Features&Author&Distance&GDP\\ \hline
      10\%      &1.0000&0.0000&0.0000\\ \hline
      20\%      &1.0000&0.0000&0.0000\\ \hline
      30\%      &1.0000&0.0000&0.0000\\ \hline
      40\%      &0.9229&0.0772&0.0000\\ \hline
      50\%      &1.0000&0.0000&0.0000\\ \hline
      60\%      &0.9116&0.0267&0.0000\\ \hline
      70\%      &0.9298&0.0251&0.0000\\ \hline
      80\%      &0.8883&0.0556&0.0559\\ \hline
      90\%      &0.8793&0.0698&0.0508\\ \hline
      100\%    &0.8206&0.0954&0.0549\\ \hline
    \end{tabular}
    \label{tab:17}
  }
\end{table*}


Figure~\ref{Figure4}b shows the feature importance ranking for ICML. We observe that the sum of authors' AIF features rank first in all the features' importance ranking list for ICML. In Table~\ref{tab:11}, we observe that author-based features' importance scores are 1 from top $10\%$ to top $40\%$ features. For using different percentages of features to train the predictive model, author-based features play a crucial role, and author-based features' importance scores are beyond 0.75. Distance-based features' importance scores are between 0.04 and 0.16 from top $50\%$ features to top $100\%$ features. GDP-based features only take effect for using top $90\%$ features and top $100\%$ features.

Figure~\ref{Figure4}c illustrates the feature importance ranking for KDD. we observe that the sum of authors' AIF features rank first in feature importance ranking list. In Table~\ref{tab:12}, we observe that author-based features' importance scores are higher than 0.8. Distance-based features' importance scores are less than 0.2, and GDP-based features' importance scores are about 0.05 from using top $80\%$ features to train the predictive model.

Figure~\ref{Figure4}d shows the feature importance ranking for MM. We observe that the sum of authors' AIF feature ranks first in feature importance ranking list excluding top $50\%$ features. In Table~\ref{tab:13}, we observe that author-based features' importance scores are beyond 0.6. Geographical distance-based features' importance scores are between 0 and 0.3. GDP-based features' importance scores are about 0.1 from top $60\%$ features.

Figure~\ref{Figure4}e illustrates the feature importance ranking for MobiCom. According to Figure~\ref{Figure4}e, we observe that authors' AIF features always rank first in feature importance ranking list. In Table~\ref{tab:14}, we observe that author-based features' importance scores are beyond 0.7. Geographical distance-based features' importance scores and GDP-based features' importance scores are less than 0.15.

Figure~\ref{Figure4}f illustrates the feature importance ranking for SIGCOMM. We observe that the sum of authors' AIF features rank first in feature importance ranking list. In Table~\ref{tab:15}, we observe that author-based features' importance scores are between 0.7 and 1. Geographical distance-based features' importance scores are less than 0.2, and GDP-based features' importance scores are less than 0.1.

Figure~\ref{Figure4}g illustrates the feature importance ranking for SIGIR. We observe that the sum of authors' AIF features rank first in feature importance ranking. In Table~\ref{tab:16}, we observe that author-based features' importance scores are higher than 0.7 for using different percentages of features to train the predictive model. geographical distance-based features' importance scores are less 0.1, and GDP-based features' importance scores are less than 0.2.

Figure~\ref{Figure4}h shows the feature importance ranking for SIGMOD. According to Figure~\ref{Figure4}h, we observe that authors' AIF features rank first in feature importance ranking. In Table~\ref{tab:17}, we observe that author-based features' importance scores are beyond 0.8. Geographical distance-based features' importance scores are less than 0.1, and GDP-based features' importance scores are about 0.05 for using top $80\%$, top $90\%$ and all the features to train the predictive model. Our experiment result shows that GDP related features are not significant comparing with others. One possible reason is that the proportion of national scientific research in GDP is very small. Another possible reason is that scientific research depending on funds exists difference in different fields.

\begin{table}[htbp]
  \renewcommand{\arraystretch}{1.2}
  \centering
  \caption{\bf NDCG@20 results for GBDT model.}
  \begin{tabular}{|c|c|} \hline
    conference&GBDT\\ \hline
    FSE       &0.604\\ \hline
    ICML      &0.851\\ \hline
    KDD       &0.909\\ \hline
    MM        &0.790\\ \hline
    MobiCom   &0.571\\ \hline
    SIGCOMM   &0.769\\ \hline
    SIGIR     &0.906\\ \hline
    SIGMOD    &0.822\\ \hline
  \end{tabular}
  \label{tab:18}
\end{table}


\begin{table*}[htbp]
  \renewcommand{\arraystretch}{1.2}
  \centering
  \caption{\bf NDCG@20 results for XGBoost model, using top percentages of features with historical relevance scores.}
  \begin{tabular}{|c|c|c|c|c|c|c|c|c|c|c|} \hline
    conference &10\%           &20\%           &30\%  &40\%  &50\%  &60\%  &70\%  &80\%  &90\%  &all features\\ \hline
    FSE        &0.564          &0.623          &0.683 &0.649 &0.661 &0.686 &0.676 &0.700 &0.696 &0.612\\ \hline
    ICML       &0.901          &0.888          &0.902 &0.915 &0.903 &0.895 &0.893 &0.892 &0.887 &\textbf{0.923}\\ \hline
    KDD        &0.937          &0.939          &0.934 &0.926 &0.930 &0.936 &0.929 &0.928 &0.928 &\textbf{0.945}\\ \hline
    MM         &0.858          &0.873          &0.823 &0.816 &0.816 &0.816 &0.854 &0.861 &0.865 &0.846\\ \hline
    MobiCom    &\textbf{0.752} &0.728          &0.642 &0.639 &0.613 &0.604 &0.600 &0.622 &0.677 &0.694\\ \hline
    SIGCOMM    &0.829          &0.839          &0.829 &0.826 &0.843 &0.840 &0.840 &0.847 &0.849 &0.803\\ \hline
    SIGIR      &0.896          &\textbf{0.922} &0.902 &0.900 &0.889 &0.889 &0.906 &0.908 &0.913 &0.912\\ \hline
    SIGMOD     &0.794          &0.779          &0.822 &0.846 &0.857 &0.854 &0.843 &0.850 &0.849 &0.839\\ \hline
  \end{tabular}
  \label{tab:19}
\end{table*}

\begin{table*}[!t]
  \renewcommand{\arraystretch}{1.2}
  \centering
  \caption{\bf NDCG@20 results for XGBoost model, using top percentages of features without historical relevance scores.}
  \begin{tabular}{|c|c|c|c|c|c|c|c|c|c|c|}                                                     \hline
    conference&10\% &20\%          &30\% &40\% &50\%          &60\% &70\%          &80\% &90\% &all features\\ \hline
    FSE       &0.712&0.680         &0.693&0.702&\textbf{0.727}&0.717&0.708         &0.716&0.704&0.714\\ \hline
    ICML      &0.857&0.860         &0.861&0.865&0.879         &0.866&0.867         &0.878&0.870&0.863\\ \hline
    KDD       &0.919&0.918         &0.929&0.935&0.920         &0.940&0.940         &0.927&0.940&0.941\\ \hline
    MM        &0.829&\textbf{0.890}&0.886&0.862&0.870         &0.847&0.850         &0.856&0.854&0.857\\ \hline
    MobiCom   &0.589&0.573         &0.615&0.652&0.633         &0.614&0.646         &0.640&0.708&0.622\\ \hline
    SIGCOMM   &0.824&0.835         &0.843&0.832&\textbf{0.852}&0.839&0.846         &0.840&0.835&0.841\\ \hline
    SIGIR     &0.889&0.881         &0.881&0.887&0.897         &0.894&0.899         &0.897&0.893&0.887\\ \hline
    SIGMOD    &0.856&0.846         &0.846&0.852&0.847         &0.851&\textbf{0.862}&0.844&0.849&0.846\\ \hline
  \end{tabular}
  \label{tab:20}
\end{table*}

An interesting finding is that we still obtain a good predictive performance despite of ignoring the historical relevance scores of institutions. The prediction accuracy NDCG@20 of GBDT model for the selected conferences is presented in Table~\ref{tab:18}. The prediction accuracy NDCG@20 of XGBoost model with historical relevance scores for the conferences is presented in Table~\ref{tab:19}. The prediction accuracy NDCG@20 of XGBoost model without historical relevance scores for the conferences is presented in Table~\ref{tab:20}. In Table~\ref{tab:18}-\ref{tab:20}, we observe that the best prediction performance of the impact of institutions is random for different top percentages of features. XGBoost model with or without relevance scores features generally performs much better than GBDT model. XGBoost model with all features including relevance scores of institutions to train data has stronger predictability for ICML, and KDD. Compared to the prediction result of GBDT with 0.851, the best prediction performance NDCG@20 with XGBoost model with all features is 0.923 for ICML. To KDD, the predictability NDCG@20 of using XGBoost model with all features is 0.945, which is higher than the prediction accuracy NDCG@20 with 0.909 of GBDT model. To FSE, the best prediction accuracy NDCG@20 with 0.727 is from XGBoost model excluding relevance scores of institutions by using top $50\%$ features. However, the prediction performance of GBDT model is only 0.604. To MM, using XGBoost with top $20\%$ features excluding historical relevance scores generates the highest prediction accuracy NDCG@20 with 0.890, While NDCG@20 of using GBDT model is only 0.790. To MobiCom, the prediction accuracy NDCG@20 of GBDT model is 0.571, while NDCG@20 of XGBoost model using top $10\%$ features to train is 0.752. the prediction performance NDCG@20 of GBDT model is 0.769 for SIGCOMM, while NDCG@20 of the XGBoost model using the top $50\%$ features excluding historical relevance scores is 0.852. To SIGIR, the prediction performance NDCG@20 of XGBoost model using top $20\%$ features is 0.922, which is higher than the prediction result of GBDT with 0.906. For SIGMOD, the prediction performance NDCG@20 of XGBoost model using top $70\%$ features excluding relevance scores of institutions is 0.862.

\section{Conclusion}
In this paper, we study a data-driven method for predicting the contributions of different institutions in eight top conferences. Previous studies have mainly focused on adopting historical relevance scores of each institution to predict the impact of institutions. Unlike previous researches, by exploring the factors that
can drive the changes of the impact of institutions such as author-based features, geographical distance-based features, economic features, and the relevance scores-based features, we have developed a high-performance prediction model, which has the ability to generate accurate predictions and explain which features have contribution to the predictive performance.

Several important findings are listed as follows: (1) the relevance scores of the accepted papers of each institution are not the most crucial factors for the prediction performance for top conferences. Via feature selection, we find that author-based features are critical in determining the number of accepted papers for an institution in the future. Compared to the $Q$ value and H-index, the AIF features are more relevant to the number of accepted papers for an institution. (2) to ICML, KDD, MobiCom, and SIGCOMM, the authors' impact such AIF, $Q$ value, and H-index are the most relevant factors for predicting the impact of institutions.
(3) for the selected top conferences excluding ICML and KDD, the performance of prediction using the fractional features is better than using all the features. (4) for KDD and SIGCOMM, the performance of prediction without using relevance scores of each institution is approximate to using the relevance scores of each institution. (5) geographic location of institution feature and state GDP feature can improve the predictive performance. Therefore, we draw a conclusion that the data-driven methods are crucial to the success of predictive models.

In the future, we will further explore the relationships between the impact of institutions and the features driving the impact of institutions change to enhance the prediction performance. In addition, this work is conducted only on literatures from the eight top conferences based on MAG dataset, examining other conferences for the same observed patterns could widen the significance of our findings.
\bibliographystyle{IEEEtran}
\bibliography{IEEEabrv}
\vfill
\end{document}